\documentclass[a4paper,12pt, epsfig]{article}
\usepackage{epsfig}
\usepackage{epstopdf}
\usepackage{graphicx}

\usepackage{amsmath}
\usepackage[psamsfonts]{amssymb}
\usepackage{euscript}

\usepackage{latexsym}

\def\hsp{,\hspace{.7cm}}

\renewcommand{\theequation}{\arabic{section}.\arabic{equation}}
\renewcommand{\(}{\begin{equation}}
\renewcommand{\)}{end{equation} \vspace{-.05in}\linebreak}

\newcounter{saveeqn}
\newcounter{savealpheqn}

\newcommand{\alpheqn}{\setcounter{saveeqn}{\value{equation}}%
  \stepcounter{saveeqn}\setcounter{equation}{0}%
  \renewcommand{\theequation}{\mbox{\arabic{section}.\arabic{saveeqn}
\alph{equation}}}
  \renewcommand{\)}{\end{equation}}}
\def\part#1{\frac{\partial}{\partial{#1}}}%
\def\group#1{\refstepcounter{equation}\setcounter{saveeqn}
 {\value{equation}}%
  \label{#1}\setcounter{equation}{0}%
\renewcommand{\theequation}{\mbox{\arabic{section}.\arabic{saveeqn}
\alph{equation}}}
  \renewcommand{\)}{\end{equation}}}
\newcommand{\reseteqn}{\setcounter{equation}{\value{saveeqn}}%
  \renewcommand{\theequation}{\arabic{section}.\arabic{equation}}%
  \renewcommand{\)}{\end{equation}}}

\newcommand{\aalpheqn}{\setcounter{saveeqn}{\value{equation}}%
  \stepcounter{saveeqn}\setcounter{equation}{0}%
  \renewcommand{\theequation}{\mbox{
        \Alph{subsection}.\arabic{saveeqn}\alph{equation}}}
   \renewcommand{\)}{\end{equation}}}
\newcommand{\areseteqn}{\setcounter{equation}{\value{saveeqn}}%
  \renewcommand{\theequation}{\Alph{subsection}.\arabic{equation}}%
  \renewcommand{\)}{\end{equation}}}

\renewcommand{\thefootnote}{\alph{footnote}}
\renewcommand{\(}{\begin{equation}}
\renewcommand{\)}{\end{equation}}
\newcommand{\ba}{\begin{eqnarray}}
\newcommand{\ea}{\end{eqnarray}}

\newcommand{\bp}{\mathop{\vtop{\ialign{##\crcr
   $\hfil\displaystyle{}\hfil$\crcr\noalign{\kern-13pt\nointerlineskip}
   \BIG{(}\hskip0pt\crcr\noalign{\kern3pt}}}}}
\newcommand{\cbp}{\mathop{\vtop{\ialign{##\crcr
   $\hfil\displaystyle{}\hfil$\crcr\noalign{\kern-13pt\nointerlineskip}
   \BIG{)}\hskip0pt\crcr\noalign{\kern3pt}}}}}
\newcommand{\pa}{\mathop{\vtop{\ialign{##\crcr
    
$\hfil\displaystyle{\oplus}\hfil$\crcr\noalign{\kern+1pt\nointerlineskip 
}
   \hspace{.08in}$^{\alpha=0}$\hskip6pt\crcr\noalign{\kern3pt}}}}}
\renewcommand{\hsp}{,\hspace{.3in}}

\newcommand{\beq}{\begin{equation}}
\newcommand{\eeq}{\end{equation}}

\numberwithin{equation}{section}

\def\hsp#1{\hspace{#1in}}

\catcode`\@=11
\def\vereq#1#2{\lower3pt\vbox{\baselineskip1.5pt \lineskip1.5pt
\ialign{$\m@th#1\hfill##\hfil$\crcr#2\crcr\sim\crcr}}}
\catcode`\@=12

\makeatletter
\newcommand\figcaption{\def\@captype{figure}\caption}
\newcommand\tabcaption{\def\@captype{table}\caption}
\makeatother
\renewcommand{\(}{\begin{equation}}
\renewcommand{\)}{\end{equation}}

\renewcommand{\beq}{\begin{equation}}
\renewcommand{\eeq}{\end{equation}}
\newcommand{\bea}{\begin{eqnarray}}
\newcommand{\eea}{\end{eqnarray}}
\newcommand{\beas}{\begin{eqnarray*}}
\newcommand{\eeas}{\end{eqnarray*}}

\newcommand{\bquo}{\begin{quote}}
\newcommand{\enqu}{\end{quote}}

\def\hsp{,\hspace{.2cm}}

\begin{document}

\begin{titlepage}

\def\thefootnote{\fnsymbol{footnote}}

\begin{center}
{\large {\bf
Calibrating Effective Ia Supernova Magnitudes\\ using the Distance Duality Relation
  } }

\bigskip

\bigskip

{\large \noindent Jarah
Evslin\footnote{\texttt{jarah@impcas.ac.cn}}}
\end{center}

\renewcommand{\thefootnote}{\arabic{footnote}}

\vskip.7cm

\begin{center}
\vspace{0em} {\em  Institute of Modern Physics, NanChangLu 509, Lanzhou 730000, China}

\vskip .4cm

\vskip .4cm

\end{center}

\vspace{1.0cm}

\noindent
\begin{center} {\bf Abstract} \end{center}

\noindent
Using only Ia supernova (SN) observations, it is not possible to distinguish the evolution of the SN absolute magnitude $M_B$ from an arbitrary evolution of the Hubble parameter $H(z)$.  However, using Etherington's distance-duality relation, which relates the angular and luminosity distances, together with the observed angular baryon acoustic oscillation (BAO) scale at any redshift $z$, one may calibrate an effective $M_B(z)$.  This calibration involves a scale which depends on the cosmological model, however the evolution of the effective $M_B(z)$ between two redshifts with BAO observations is independent of this scale.  The line of sight BAO scale can be used to extend this calibration to redshifts near $z$.  As an application, using BOSS BAO at $z=0.32$ and $2.34$, JLA supernova at low $z$ and Hubble Space Telescope SN at $z>1.7$, we find a statistically insignificant downward shift $M_B(2.34)-M_B(0.32)=-0.08\pm 0.15.$   Replacing BOSS data with the best fit Planck $\Lambda$CDM BAO expectations, we find a shift of $-0.24\pm 0.13$.  With the SN that will be observed by the James Webb Space Telescope, such a calibration at $z=2.34$ will be more precise, and it will serve as an anchor for cosmological analyses with the SN that it will observe at yet higher $z$.

\vfill

\begin{flushleft}
{\today}
\end{flushleft}
\end{titlepage}

\section{Introduction}

High redshift type Ia supernova (SN) tend to be redder \cite{union2} and have higher mass, lower metallicity progenitors than those at low redshifts.  Thus one expects that their absolute magnitudes $M_B(z)$ will depend upon the redshift $z$, even after color and spectral shape corrections \cite{dominguez}.  While corrections for color, shape and host galaxy mass are standard practice for the derivation of the Hubble diagram from SN observations, the derivation of the Hubble parameter $H(z)$ requires that this corrected $M_B(z)$ be independent of $z$.  Therefore the $z$-evolution of the corrected $M_B(z)$ is a source of systematic error for the use of SN to determine the Hubble diagram.

There is a vast literature on the calibration of $M_B(z)$ using independent determinations of the luminosity distance, determined via the angular diameter distance which may be measured using standard rulers such as BAO \cite{aubourg} or standard sirens such as gravity waves from inspiraling binaries.  Below we will follow a similar strategy, however with two novelties.

First, we will use the fact that the BAO ruler is independent of redshift but will not use any calibration of its length.  In light of the current 9\% (more than $3\sigma$) tension between between the local distance ladder \cite{riessladder} and early time cosmological measurements of present-day expansion, an analysis which is independent of the absolute distance scale seems prudent.  The use of relative distances means that we cannot calibrate the absolute magnitude at any redshift, but rather we can obtain its evolution.  This evolution in fact is all that one needs to use high redshift SN for cosmology.

Second, while applications of such calibrations to date have used only BAO measurements at redshifts $z\lesssim 1$, we will apply this technique to a BAO measurement at $z=2.34$.  However, as has been repeatedly emphasized in Ref.~\cite{hoflich}, at $z>0.7$ the age of the universe is less than or equal to the expected progenitor age, and so one expects significant supernova evolution.  Indeed, a steep drop in the color correction at $z\sim 0.7$ has been reported in Ref.~\cite{shariff}.

A comparison of SN and BAO at such high redshifts would not have been possible even a year ago, it is possible now for the first time because of the high redshift SNe very recently discovered by the CANDELS survey.  However this extension to high $z$ comes at a high price.  The supernova and BAO redshifts are not quite the same, and so a somewhat unappealing redshift extrapolation will be necessary.  We will apply generous error bars to this extrapolation.  In fact the dominant uncertainties arise not from the extrapolation, but from the uncertainties in the SN observations themselves.  These uncertainties are so large that in our opinion a sophisticated analysis is not warranted at this time, the analysis presented below is intended to be crude but as independent of astrophysical assumptions as possible.

There has been much discussion following the high redshift BAO observations of possible new systematics in BAO measurements.  We have nothing to add to this discussion, and so we will simply present two analyses, one which uses the BAO uncertainties as reported by BOSS and another which uses not the observed BAO, but rather the BAO scale that would be expected in a Planck best fit $\Lambda$CDM cosmology.

Before beginning, some clarification is warranted regarding our definition of the corrected absolute magnitude $M_B(z)$, whose evolution is to be determined below.  It is not our goal to characterize the astrophysical evolution of SN.  Our goal is rather to characterize a systematic in a standard SN analysis using the SALT2 light curve fitter and $K$-corrections determined with the usual precision.  In particular, we will not attempt to distinguish observation-related errors, fitting errors and intrinsic evolution.  Therefore our quantity $M_B(z)$ is defined to be the corrected absolute magnitude which would be determined from standard observations and analyses, and so it already includes corrections for grey dust, $K$-corrections and $S$-corrections.  It also assumes the standard color, shape and host light curve corrections, which in our analysis we will set to their JLA values although this last assumption may be removed when more high redshift data is available.  More sophisticated analyses using more information about each SN, such as more bands and more complicated shape analyses, would yield a different corrected $M_B(z)$.  An application of the methodology below to such cases would require a new calculation but would be straightforward.

This approach has the disadvantage that it yields a quantity which cannot be directly compared with SN simulations.  However it has the advantage that it is the quantity which must be understood if one wishes to use higher $z$ supernova, for example those that will be discovered by the James Webb Space Telescope, to characterize the expansion of the universe.

\section{Calibration using SN Observations Alone?}

It is often claimed \cite{livio,salzano} that the evolution can be quantified by observing more SN.  However such studies assume certain forms of the dark energy equation of state \cite{salzano}, or at least that the dark energy density becomes small at redshifts of $2$ to $4$ \cite{livio}.  These papers are often misconstrued to simply imply that future SN surveys will be able to determine the evolution of the corrected absolute magnitudes of SNe \cite{rodneysn}.

However, if no restriction is placed on the dark energy equation of state, or more generally on the dark energy density as a function of redshift, then the Hubble parameter $H(z)$ may be an arbitrary function.  Any $z$-dependence in the corrected SN absolute magnitude $M_B(z)$ may be separated from $M_B(z)$ by defining
\beq
f(z)=M_B(z)-M_B(0).
\eeq
If $d_L(z)$ is the luminosity distance to $z$ and $m_B^\prime(z)$ is the average color, shape and host-corrected observed magnitude at redshift $z$ then
\bea
d_L(z)&=&10^{\frac{m_B^\prime (z)-M_B(z)}{5}+1}{\rm{pc}}\label{dl}\\
&=&10^{-\frac{f(z)}{5}}e^{\frac{m_B^\prime (z)-M_B(0)}{5}+1}{\rm{pc}}.
\eea
Therefore any $z$-dependent $M_B$ can be consistently interpreted as a cosmology in which $d_L$ is set to
\beq
d_L^\prime(z)=10^{\frac{f(z)}{5}} d_L(z)
\eeq
and $M_B(z)=M_B(0)$ is constant. The corresponding Hubble parameter $H^\prime(z)$ can be found by integrating the defining equation
\beq
d_L^\prime(z)=(1+z)\int_0^z\frac{c dz^\prime}{H^\prime(z^\prime)}.
\eeq
In particular, for any $z$-dependent function $M_B(z)$, a solution $H^\prime(z)$ exists
\beq
H^\prime(z)=\frac{c}{\partial_z\left(d_L^\prime(z)/(1+z)\right)}.
\eeq
Therefore by approximating $M_B(z)$ to be a constant $M_B(0)$, one concludes that the Hubble parameter is $H^\prime(z)$ and not the true $H(z)$.  More measurements simply improve the precision with which $f(z)$ is known, rather than reducing $f(z)$, and so do not allow one to distinguish $H(z)$ from $H^\prime(z)$ unless a constraint is placed on the form of $H(z)$.  However with all of the dark energy models which are available today \cite{horndeski,braiding}, it is difficult to physically motivate such a constraint without the use of auxiliary data sets.  We thus conclude that auxiliary datasets are the only reliable way to break the perfect degeneracy between the evolutions of $H(z)$ and $M_B(z)$.

\section{Anchoring SN and BAO}

The baryon acoustic oscillation (BAO) feature is a narrow peak in the position-space matter-matter correlation function.  Its location, $r_d$, is fixed in comoving coordinates and so it provides a standard ruler at all redshifts since the drag epoch \cite{baoteor}.

The BAO scale may be measured along the line of sight direction, as a correlation between objects with redshifts separated by $\Delta z(z)$, or else perpendicular to the line of sight direction, where it represents a correlation between objects separated by an angle $\Delta\theta(z)$.  These two observables each yield a distance measure to a given redshift
\beq
\Delta z(z)=\frac{r_d}{d_H(z)}\hsp
\Delta\theta(z)=\frac{r_d}{d_A(z)}. 
\eeq
In a flat, FLRW cosmology these distances can be expressed in terms of the Hubble parameter~$H(z)$
\beq
d_H(z)=\frac{c}{H(z)}\hsp
d_A(z)=\frac{c}{1+z}\int_0^z\frac{dz^\prime}{H(z^\prime)}. \label{frw}
\eeq
Our method for calibrating the SN magnitude in principle does not require the flat, FLRW assumption (\ref{frw}).  However we will use this assumption below to increase the number of SN which can be used in this calculation, increasing the precision obtained.

In 1933, Etherington obtained a powerful relation between the angular diameter distance $d_A(z)$, which is measured using angular BAO, and the luminosity distance $d_L(z)$, which is measured using SN.  This relation is called the distance-duality relation \cite{etherington}
\beq
d_L(z)=(1+z)^2d_A(z) \label{dd}
\eeq
where $z$ is the Doppler shift between the two objects, which coincides with the redshift for locally comoving objects in an FLRW universe.  As he wrote, it requires only that light travel along null geodesics, for example if light is massless and minimally coupled to a metric.  Ellis {\it{et. al.}} have used the perfect blackbody normalization of the cosmic microwave background radiation measured by COBE to argue that the distance-duality relation is respected by our cosmology with a precision of $10^{-4}$ at all redshifts up to recombination~\cite{ellis}.  Therefore in what follows we will simply assume that the distance-duality relation is an exact equality.

It is conventional to add the assumption that there be no grey dust, or in other words that $d_L(z)$ be calculated assuming that no light is absorbed before its arrival, for example by correcting for absorption by dust which has been estimated using spectral distortions.  We believe that a sizable grey dust contribution is very difficult to reconcile with the aforementioned precision COBE measurement of the normalization of the CMB spectrum, which excludes even a $10^{-4}$ loss of luminosity.  To evade this bound, grey dust needs to be not quite grey and it can only affect the wavelengths relevant for SN observations at $z<<100$.

For our present application, we do not need to assume the absence of grey dust.  Dust decreases the observed light, and so increases $m_B^\prime(z)$ while keeping $d_L(z)$ fixed by the distance-duality relation.  Then Eq.~(\ref{dl}) implies that our $M_B(z)$ increases so as to keep $m_B^\prime(z)-M_B(z)$ constant.  Therefore, if there is indeed grey dust, the formulae below may nonetheless be applied but it must be understood that $M_B(z)$ is the dust-corrected absolute magnitude, and so is in general higher than the true magnitude.  This distinction is irrelevant for using SN as standard candles to map the Hubble diagram, but of course it means that one can no longer match $M_B(z)$ with the predictions of stellar evolution models, which will not have the grey dust correction.

Note that this argument continues to hold if the grey dust has an arbitrary $z$ dependence, so long as it is isotropic.  Whatever the $z$ dependence of the grey dust, it will simply correct the $z$ dependence of the effective absolute magnitude $M_B(z)$.  However it is precisely the same effective absolute magnitude whose evolution we calibrate using BAO, and it is the effective absolute magnitude which, at other values of $z$, is used to determine the luminosity distance and so the cosmological evolution.  Therefore even if the grey dust evolves with $z$, as it does in most models, this does not inhibit the mapping of the Hubble diagram.  Of course implicit in this approach is the assumption that the effective $M_B(z)$ is sufficiently well-behaved that a calibration at some value of $z$, say $z=2.3$, yields information about the value at a nearby value of $z$, say $z=3$.  This assumption is satisfied in the supernova luminosity evolution studies and grey dust models of which we are aware, in which both effects are always monotonic as the former is generally tied to the mean progenitor metallicity and the latter effect increases with distance.

The distance-duality relation (\ref{dd}) between the derived distance scales $d_A$ and $d_L$ implies a relation between the corresponding observables, obtained from Eq.~(\ref{dl})
\beq
\frac{d_A(z)}{r_d}=\frac{{\rm{pc}}}{(1+z)^2A(z)}10^{\frac{m_B^\prime(z)}{5}} \label{baosn}
\eeq
where we have defined the {\it{anchor}}
\beq
A(z)=r_d 10^{\frac{M_B(z)}{5}-1}. 
\eeq
The anchor $A(z)$ is the distance at which a magnitude 5 star has the same apparent magnitude as a magnitude $M_B(z)$ supernova at a distance of $r_d$.  If $M_B(z)$ is $z$-independent, then so is the anchor.

One may use Eq.~(\ref{baosn}) to calculate $A(z)$ at redshifts where SN and BAO data are available, allowing a calibration of SN from BAO \cite{aubourg} or vice versa.  If the values of $A(z)$ are inconsistent with one another, this implies that $M_B(z)$ evolves in time.

Unfortunately $A(z)$ depends on $r_d$, which depends on the cosmological model, for example the number of neutrino flavors, and also is not directly observable.  However ratios of $A(z)$ at distinct redshifts are independent of $r_d$ and so can yield model-independent differences in $M_B(z)$.  In particular if $A(z)$ is determined at redshifts $z_1$ and $z_2$ from angular BAO and SN measurements combined via Eq.~(\ref{baosn}), then
\bea
&&M_B(z_2)-M_B(z_1)=5{\rm{Log}}_{10}\left(\frac{A(z_2)}{A(z_1)}\right) \label{princ}\\
&=&5{\rm{Log}}_{10}\left(\left(\frac{1+z_1}{1+z_2}\right)^2\left(\frac{d_A(z_1)}{d_A(z_2)}\right)10^{\frac{m_B^\prime(z_2)-m_B^\prime(z_1)}{5}}\right).\nonumber
\eea
This is our main result.  Note that the ratio of angular diameter distances is observable with BAO, without knowing $r_d$, as it is simply $\Delta\theta(z_2)/\Delta\theta(z_1)$.

\section{Extrapolating SN Data to Nearby ${\mathbf{z}}$}

How does one determine the corrected magnitudes $m_B^\prime(z)$ from the SN data? For sufficiently large SN data sets, the necessary extrapolation needs to be done carefully as a result of systematic errors relating data at distinct redshifts, nonetheless various frequentist \cite{jla} and Bayesian \cite{mastat} methods exist.

The situation is more subtle for redshifts with sparse SN data, such as $z=2.34$ where BOSS measured the BAO scale in the Lyman $\alpha$ forest \cite{1404.1801,fontribera}.  In such a case, SN lie necessarily at distinct redshifts from the BAO data, and so the SN data must be extrapolated to distinct but nearby redshifts.

By combining Eq. (\ref{frw}), which is valid in an FLRW cosmology, with Eq.~(\ref{dd}) and differentiating with respect to the redshift we obtain the simple identity
\beq
\frac{\partial d_L(z)}{\partial z}=(1+z)\left(d_H(z)+d_A(z)\right).
\eeq
We extrapolate SN data by integrating this equation, with the crude approximation that $d_H(z)$ and $d_A(z)$ are constant over the range of integration.  While the evolution of $d_H$ and $d_A$ are in opposite directions and so somewhat cancel one another, and while dark energy lessens the evolution, we conservatively set the uncertainty in this extrapolation to be the entire expected evolution of $d_H$ alone in a universe with only dark matter.  Even with this large estimate for the error, when extrapolating SN data  at $z>1.7$ the extrapolation uncertainties are subdominant to the SN measurement uncertainties.

Expressed in terms of the corrected magnitudes, and approximating the anchor $A(z)$ or equivalently $M_B(z)$ to be $z$-independent over the range of redshifts considered, this extrapolation is then
\bea
10^{\frac{m_B^\prime(z_2)}{5}}&=&10^{\frac{m_B^\prime(z_1)}{5}}+\left((z_2-z_1)+\frac{z_2^2-z_1^2}{2}\right)\frac{A}{\rm{pc}}\nonumber\\&&\times\left(\frac{d_H(z_2)}{r_d}+\frac{d_A(z_2)}{r_d}\right). \label{extra}
\eea
Notice that the extrapolation uses both the measured angular and line of sight BAO scales.

\section{Data Sets and Results}

\subsection{Data Sets}

We use the BOSS angular BAO scales measured at effective redshifts of $z=0.32$, $z=0.57$ whose analyses in Refs.~\cite{1509.06371} and \cite{1509.06373} were combined in Ref.~\cite{1509.06373}.  We also use the BOSS angular and line of sight BAO scales measured at $z=2.34$ in the Lyman $\alpha$ forest autocorrelation function in Ref.~\cite{ffbao} and in the forest-quasar correlation function in \cite{fqbao}.  These Lyman $\alpha$ forest results were combined in Ref.~\cite{ffbao}.  All BAO data used in this note is summarized in Table~\ref{baotab}.

\begin{table}
\centering
\begin{tabular}{c|l|l|l|}
&$z=0.32$&$z=0.57$&$z=2.34$\\
\hline\hline
$d_A(z)/r_d$&$6.76\pm 0.15$&$9.47\pm 0.13$&$10.93\pm 0.35$\\
\hline
$d_H(z)/r_d$&not used&not used&$9.15\pm 0.2$\\
\hline
\end{tabular}
\caption{BAO measurements used in this note}
\label{baotab}
\end{table}

To use Eq.~(\ref{princ}), we need SN data at the same redshifts.  At the lower two redshifts, we use the JLA data set and analysis presented in Ref.~\cite{jla}.  In this paper, the corrected magnitude $m_B^\prime$ of an SN is related to the observed magnitude $m_B$ by the relation
\beq
m_B^\prime=m_B+\alpha x - \beta c - \delta \label{mb}
\eeq
where $x$ and $c$ are real numbers describing the shape and color of the SN light curve, obtained using the SALT2 light curve fitter \cite{salt2}.  $\delta$ is a correction for the host galaxy mass, which is set to zero if the mass is less than $10^{10}M_\odot$ and otherwise to a constant.  This constant, along with $\alpha$ and $\beta$, were determined in \cite{jla} by fitting the data to a cosmological model.  In general, the values of $\alpha$, $\beta$ and $\delta$ will have little dependence on the model if there is ample data at fixed redshift.  However, incorporating the $\alpha$ and $\beta$ dependence of the variance into this fit requires some arbitrary choices, and those of \cite{jla} in a Bayesian interpretation would correspond to an unusual prior \cite{mastat}.  As they have little effect on our results, we will simply adopt the best fit values from Ref.~\cite{jla}
\beq
\alpha=0.141\pm 0.006\hsp
\beta=3.101\pm 0.75\hsp
\delta=-.07. \label{jlaparam}
\eeq

We then read the observed SN magnitudes from the bottom panel of their Fig. 11, using an average of the SDSS and SNLS values in the corresponding bin to obtain the correction to the best fit to their best fit cosmology.  We find $10^{m_B^\prime(0.32)/5}=(2.59\pm 0.03)\times 10^4$ and $10^{m_B^\prime(0.57)/5}=(5.15\pm 0.10)\times 10^4$.  Then the BAO-SN anchor can be found using Eq.~(\ref{baosn})
\beq
A(0.32)=2.18\pm 0.05\ \rm{kpc},\ 
A(0.57)=2.21\pm 0.04\ \rm{kpc}.
\eeq
The consistency of these anchors, noted also in Ref.~\cite{congma}, bounds the evolution of the corrected absolute magnitudes of SN between $z=0.32$ and $z=0.57$.

\begin{table}
\centering
\begin{tabular}{c|l|l|l|l}
&SCP-0401&GNS13Sto&UDS10Wil&GNS12Col\\
\hline\hline
$z$&$1.713$&$1.80$&$1.914$&$2.26$\\
\hline
$m_B$&$26.14$&$26.14\pm 0.07$&$26.20\pm 0.11$&$26.80\pm 0.07$\\
\hline
$x$&$0.2$&$-0.47\pm 0.68$&$-1.50\pm 0.51$&$0.15\pm 1.06$\\
\hline
$c$&$-0.10$&$-0.02\pm 0.07$&$-0.07\pm 0.11$&$0.04\pm 0.13$\\
\hline
$m_B^\prime$&$26.41\pm 0.15$&$26.14\pm 0.25$&$26.22\pm 0.37$&$27.05\pm 0.44$\\
\hline
ext&$2.75\pm 0.23$&$2.42\pm 0.25$&$2.34\pm 0.35$&$2.68\pm 0.61$\\
\hline
\end{tabular}
\caption{High redshift SN.  The last row lists the extrapolated values of $10^{m_B^\prime(2.34)/5}$ in units of $10^5$, including lensing and intrinsic scatter in the uncertainties.}
\label{sntab}
\end{table}


We will use four high $z$ SN discovered with the Hubble Space Telescope (HST).  SN SCP-0401 was discovered with the HST ACS instrument \cite{acssn}.  The other three, found by the CANDELS survey, are SN UDS10Wil \cite{jonessn}, SN GND13Sto and SN GND12Col \cite{rodneysn}.  The properties of these SN, as extracted with SALT2, are summarized in Table~\ref{sntab}.  Eq.~(\ref{mb}) was then used, with the best fit JLA parameters (\ref{jlaparam}), to calculate $m_B^\prime$ in each case.  The massive host correction was applied to SCP-0401.

\subsection{Malmquist Bias}
We have attempted to correct for Malmquist bias for the 3 highest redshift supernovae, which were all discovered by the CANDELS survey.  To do this, we use Fig.~15 of Ref.~\cite{rodneysn}, which reports the region where the CANDELS survey loses sensitivity to SNe with average fiducial shapes and colors.  The survey has an average cadence of about 50 days.  Two curves are reported, one at which the average supernova would be visible for exactly 50 days and one at which it would never be visible.

We have linearly interpolated between these two curves to estimate the probability that a SN with a given magnitude and redshift would be discovered.  Then at each redshift, for each of 100 equally spaced chosen magnitudes $m_i$, we have generated $10^5$ SN with magnitudes scattered about the chosen magnitudes with a Gaussian scattering equal to the total dispersion of the observed SN.  This total dispersion includes measurement uncertainties, parameter uncertainties, lensing and the intrinsic dispersion of the SN.  Then we have calculated a weighted average $m_{\rm av}$ of the observed magnitudes of the sample where each SN is weighted by its probability of being observed.  At each chosen magnitude $m_i$, we identify $m_i-m_{\rm av}$ with the Malmquist bias.

\begin{figure} 
\begin{center}
\includegraphics[width=3.2in,height=1.5in]{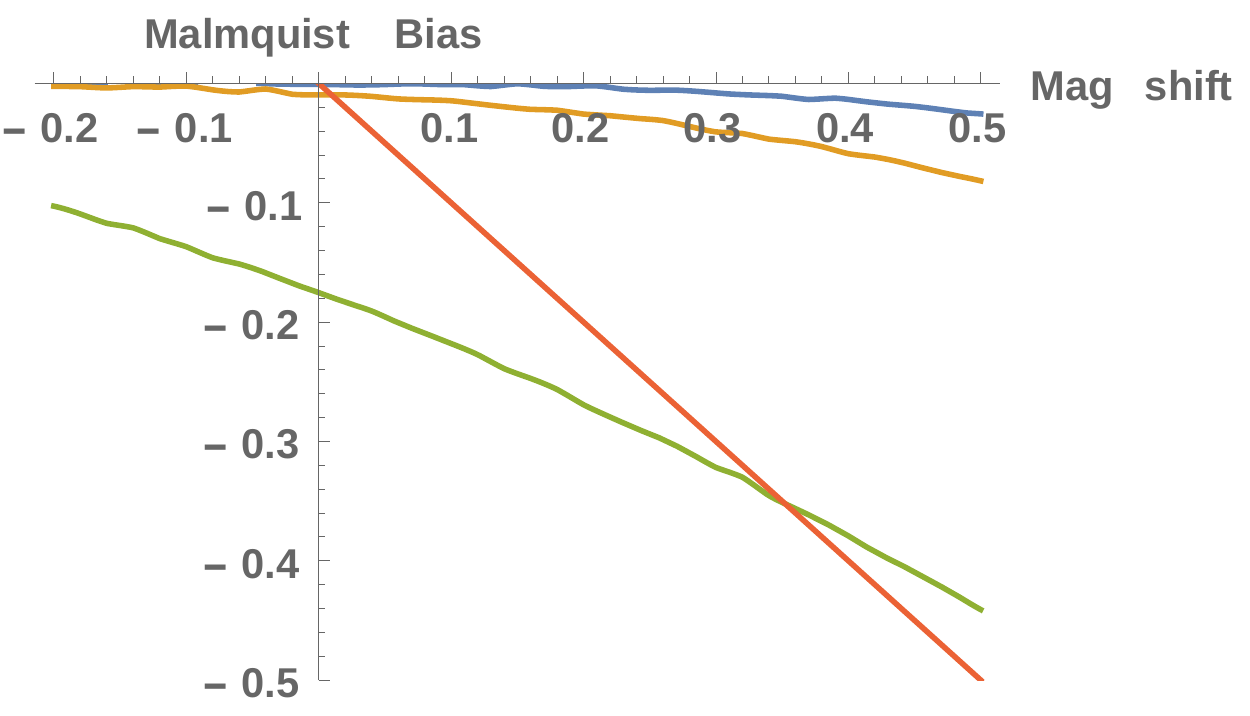}
\caption{The Malmquist bias (vertical axis) that would result from a shift in magnitude given by the horizontal axis.  The three observed SNe correspond to the three curves, from top to bottom in order of increasing redshift.  The diagonal line is the line at which the corresponding upward shift in magnitude exactly cancels the downward shift resulting from the Malmquist bias, and so its intersection with the 3 curves yields the 3 Malmquist corrections.} 
\label{malmfig}
\end{center}
\end{figure}

Our results for the three CANDELS SN are shown in Fig.~\ref{malmfig}. The horizontal axis is the difference between the chosen magnitude $m_i$ and the corrected magnitude $m^\prime_B$ of the SN actually observed by CANDELS.  The vertical axis is the Malmquist bias $m_i-m_{\rm av}$ at that magnitude.  The diagonal line represents the increase in the true magnitude which would be precisely canceled by the Malmquist bias, in other words $m^\prime_B=m_{\rm av}$.  Therefore, to correct for the Malmquist bias, we have increased the magnitude of each SN by the value of the intersection of its corresponding curve with the diagonal line.  This intersection is the upward shift in magnitude which cancels the corresponding Malmquist bias.  The highest redshift SN, GNS12Col, requires a Malmquist correction of $0.35$, the next UDS10Wil has a correction of only $0.01$ while the correction is negligible for GNS13Sto.

\subsection{Results}

\begin{figure} 
\begin{center}
\includegraphics[width=3.2in,height=1.5in]{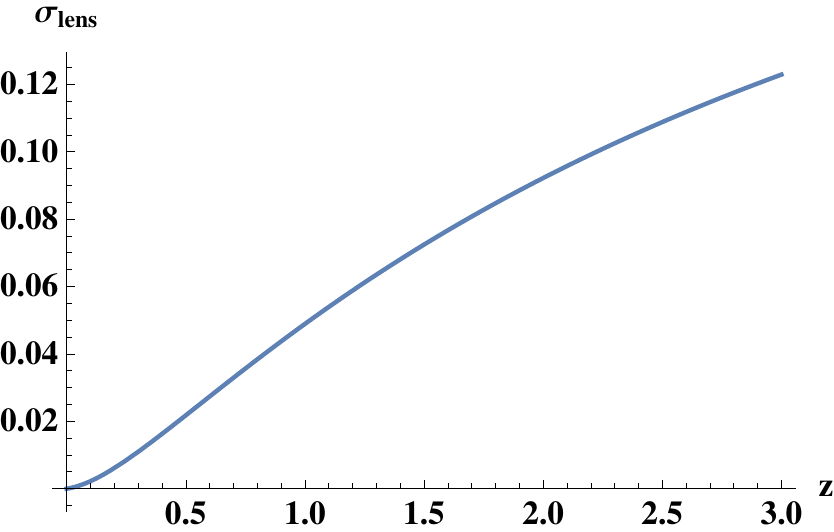}
\caption{The 1$\sigma$ variation arising from gravitational lensing as a function of redshift, as computed in Ref.~\cite{marra}.} 
\label{lensfig}
\end{center}
\end{figure}

The uncertainty on $m_B^\prime$ for SCP-0401 was fixed as in Ref.~\cite{acssn}, the uncertainties on the others were calculated by propagating the uncertainties in the parameters.  Following the estimate in Ref.~\cite{jla}, the uncertainty on each $m_B^\prime$ was added in quadrature to the expected 10.6\% magnitude scatter and, following \cite{marra}, an additional scatter was added to the uncertainty to account for lensing.  More precisely, we approximated lensing to be a Gaussian approximated with a standard deviation given by $\sigma_{\rm lens}$ from \cite{marra} which was evaluated using the Planck $\Lambda$CDM best fit parameters
\beq
\sigma_8=0.815\hsp \Omega_m=0.3121.
\eeq
The size $\sigma_{\rm lens}$ of the lensing contribution to the uncertainty as a function of redshift $z$ is shown in Fig.~\ref{lensfig}.

Then Eq.~(\ref{extra}) was applied to extrapolate the magnitudes that would have been expected at $z=2.34$.  The average of these values of $m_B^\prime(2.34)$, weighted by their uncertainties, was then substituted into Eq.~(\ref{baosn}) to obtain the anchor
\beq
A(2.34)=2.10\pm 0.14\ {\rm kpc}.
\eeq
Then Eq.~(\ref{princ}) yields our final result
\beq
M(2.34)-M(0.32)=-0.08\pm 0.15. \label{finale}
\eeq
Without the Malquist correction we would instead find an evolution of $-0.11\pm 0.15$.

\section{Variations}
\subsection{Redshift-Dependent Color and Shape Corrections}

In SN model calculations, often luminosity is correlated with metallicity which is correlated with color.  As metallicity decreases at high redshift, one thus expects a different distribution of progenitor metallicities and so colors and luminosities.  This generally implies that the optimal color correction $\beta$ will be $z$ dependent.

There have been a number of searches for such a $z$-dependence in $\beta$.  Some authors \cite{jla} found no evidence for $z$-dependence, while some found that $\beta$ increases \cite{wangwang} or decreases \cite{mohlabeng,shariff} at large $z$.  The reported evolution of $\beta$ is quite substantial, ranging from about 1 \cite{shariff} to 4 \cite{wangwang} as $z$ increases from $0$ to $1$.  In general, the evolution of $\alpha$ is not observed \cite{wangwang} or is observed only at low confidence \cite{mohlabeng}.

This suggests that the optimal color correction $\beta$ which should be applied to the four high $z$ SN analyzed in this paper may be appreciably different from that found by JLA at lower values of $z$.  Therefore, we have repeated our calculations for 100 values of $\alpha$ and $\beta$ between $-1$ and $1$ and between $-5$ and $5$ respectively.

\begin{figure} 
\begin{center}
\includegraphics[width=3.2in,height=1.5in]{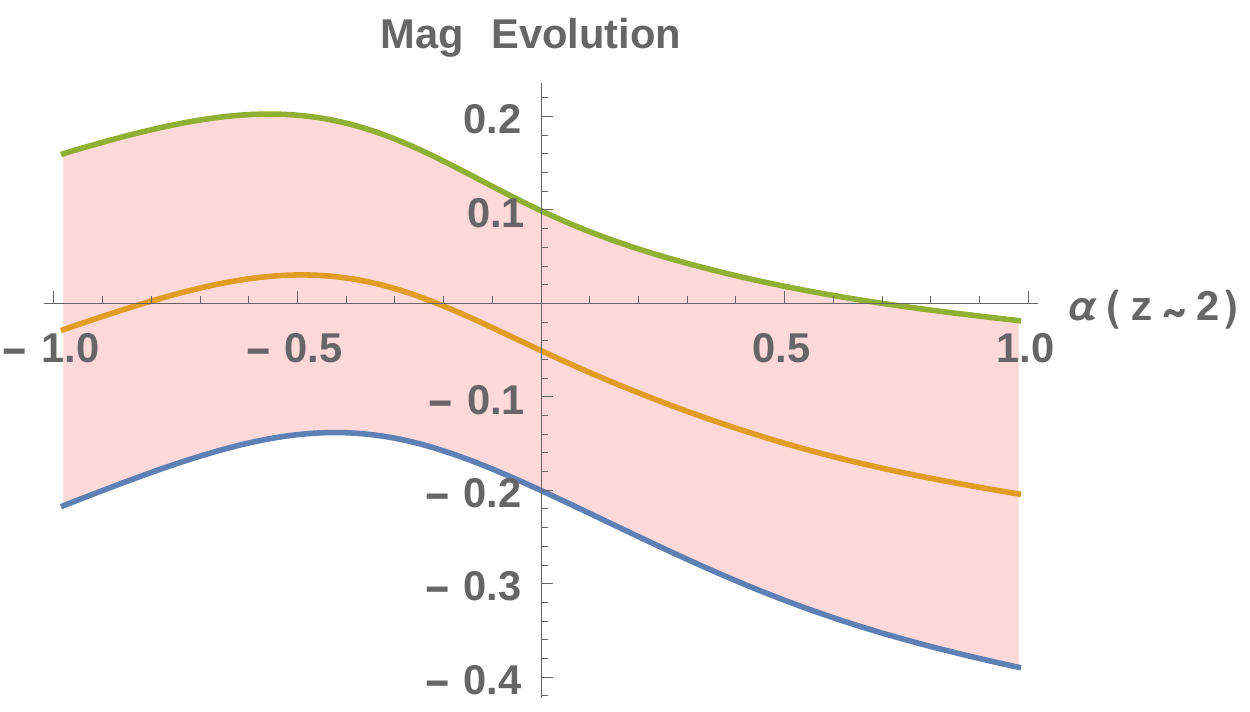}
\includegraphics[width=3.2in,height=1.5in]{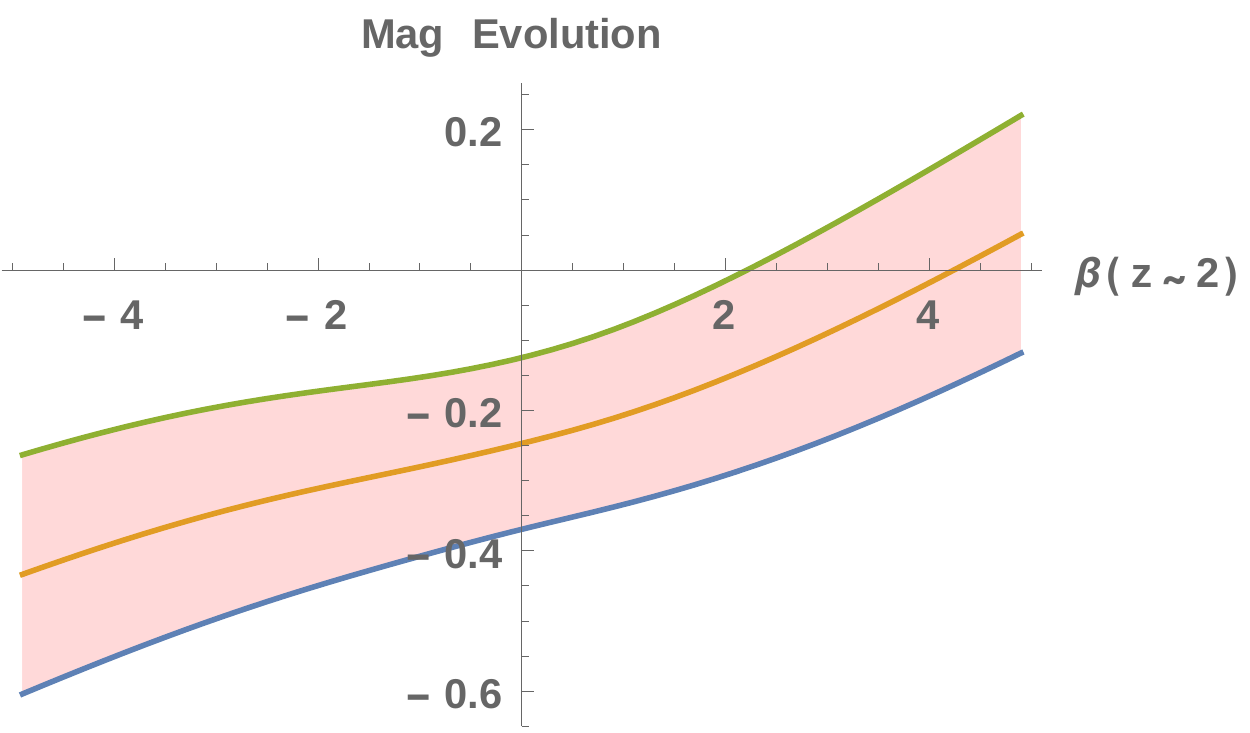}
\caption{The 1$\sigma$ allowed evolution of the effective absolute magnitude if the 4 high redshift supernova magnitudes are corrected not with the JLA parameters, but two JLA parameters and either $\alpha$ (left panel) or $\beta$ (right panel) assuming the value shown on the $x$-axis.} 
\label{abfig}
\end{center}
\end{figure}

The resulting $1\sigma$ allowed evolution of the effective absolute magnitude between $z=0.32$ and $z=2.34$ is shown in Fig.~\ref{abfig}.  While the downward evolution of $\beta$ at higher $z$ reported in Refs.~\cite{mohlabeng,shariff} increases the evidence for evolution of the effective magnitude, nonetheless 2$\sigma$ of evidence would require that $\beta(2)$ is about zero.  However, when these uncertainties are reduced by new BAO and SN datasets, even a mild reduction in $\beta$ at high redshift may result in significant evidence for SN brightening as $z$ increases.

\subsection{Replacing BAO Data with Planck BAO Predictions}

The Lyman $\alpha$ forest measurement of the BAO scale at $z=2.34$ in Ref.~\cite{1404.1801} is in tension with a $\Lambda$CDM cosmology using the best fit Planck parameters from Ref.~\cite{planck}.  Various studies \cite{1404.1801,aubourg,sahni} place the tension between 2$\sigma$ and 3$\sigma$.  This tension is particularly intriguing because, if the central value of the BAO measurement is confirmed, it will imply an evolution of the dark energy equation of state \cite{mebao1,mebao2}.

The anomaly is not apparent in the angle-averaged BAO scale.  Rather it is manifested as a 7\% deficit in $d_A$ and a 7\% excess in $d_H$ with respect to the best fit Planck $\Lambda$CDM expectations \cite{planck}.  How is the result Eq.~(\ref{finale}) affected by this anomalous BAO measurement?  We have repeated our calculation, replacing all BAO measurements with the BAO values which would be obtained in a Planck $\Lambda$CDM cosmology \cite{planck}
\beq
\Omega_m=0.3121\pm 0.0087 \hsp \frac{H_0 r_d}{c}=0.0332\pm 0.0004
\eeq
which yields BAO angular diameter measurements of
\beq
\frac{d_A}{r_d}(0.32)=6.71\pm 0.09\hsp
\frac{d_A}{r_d}(0.57)=9.38\pm 0.13\hsp
\frac{d_A}{r_d}(2.34)=11.68\pm 0.18.
\eeq
To extrapolate the supernova redshifts to $z=2.34$ we also need the Planck $\Lambda$CDM extrapolated radial scale
\beq
\frac{d_H}{r_d}(2.34)=8.55\pm 0.16.
\eeq
Repeating the above analysis with these predictions replacing the BAO data we found
\beq
M(2.34)-M(0.32)=-0.24\pm 0.13.
\eeq
In other words, if the Lyman $\alpha$ BAO measurement is incorrect due to a systematic error or an unusually large statistical fluctuation, with the true value agreeing with the Planck $\Lambda$CDM cosmology, then there would be some evidence for SN magnitude evolution.  In particular this implies that the BOSS Lyman $\alpha$ anomaly {\it{improves}} agreement with the thesis that effective supernova magnitudes do not evolve up to $z\sim 2$.

\subsection{Different Lensing Corrections}

In this paper we have used the weak lensing scatter correction to the SN magnitudes from Ref.~\cite{marra}.  However many studies in the literature use the older and larger scatter
\beq
\sigma_{\rm lens}=0.088z
\eeq
from Ref.~\cite{lens}, although it is marginally less consistent with observations \cite{lensobs}.  We have repeated our calculations with this larger lensing estimate.  In this case, using BAO data from BOSS, as summarized in Table~\ref{baotab}, we obtain
\beq
M(2.34)-M(0.32)=-0.09\pm 0.16
\eeq
while using the Planck best fit BAO predictions we obtain
\beq
M(2.34)-M(0.32)=-0.24\pm 0.14.
\eeq
Therefore at this point the choice of lensing scatter model makes little difference, uncertainties are still dominated by the small SN statistics at high $z$ and the SN measurement uncertainties.  

\section{Comparison with the Literature}

The use of the Distance-Duality relation to compare SN standard candle data with standard rulers is ever more common.  In this section we will describe how our approach differs from some of the others that have appeared so far.

Most of the literature is devoted to testing the distance-duality relation.  As the precision with which the distance-duality relation has been tested in Ref.~\cite{ellis} exceeds the precision of essentially any cosmological probe, that is not our approach.  However the calculations themselves are independent of the aim of the work and so our calculations resemble those which have appeared in several other papers.

Most recent work testing the distance-duality relation has focused on observations of clusters, comparing for example Sunyaev-Zeldovich and X-ray observations \cite{uzan,bernardis}.  This strategy requires several strong assumptions, for example Ref.~\cite{bernardis} assumes the $\Lambda$CDM cosmological model.  More recent studies such as Refs.~ \cite{holanda,santos} are independent of such cosmological assumptions.  However they require astrophysical assumptions regarding the cluster geometry.  For example Ref.~\cite{holanda} finds that an isothermal elliptic geometry is more consistent with the distance-duality relation than an isothermal spherical model, while Ref.~\cite{santos} uses nonisothermal spherical models without relaxing the spherical symmetry.

Many other papers have compared SN directly with BAO, as has been done here.  However accurate measurements of $d_A$ have become available only very recently.  Therefore, most of the older literature \cite{lazkoz,heavens} uses $d_V$ as a proxy.  The distance $d_V$ is a weighted geometric mean of $d_A$ and $d_H$ which has the advantage that, at least in the case of galaxy surveys, it can be measured more precisely than $d_A$, in particular with a small dataset.  However to use the distance-duality relation one then needs to obtain $d_A$ from $d_V$.  They are related by a differential equation which contains the unknown function $w(z)$, the dark energy equation of state.  Therefore to obtain $d_A$, strong assumptions were always necessary, such as a linearized parameterization of $w(z)$ \cite{heavens}.

A few of the more recent papers \cite{wu,congma} have used direct BAO measurements of $d_A$.  However, unlike the current paper, they fixed the BAO scale using a cosmological model.  This cosmological model explicitly contains the number of neutrino flavors and their masses.  Perhaps more seriously it relies upon interpretations of CMB data which heavily rely on the assumption that dark energy results entirely from a cosmological constant.  This is not problematic for the stated goal of those papers, a test of the distance-duality relation.  However our goal is to provide a tool to allow SN data to determine the evolution of dark energy or equivalently the cosmological expansion.  Therefore, in the present paper, as in Ref.~\cite{heavens}, the BAO standard ruler remains uncalibrated. 

In summary, to our knowledge all previous papers combining BAO, SNe and the distance-duality relation did one of the following.  Some used only the isotropic BAO measurements of $d_V$ and then assumed a particular class of dark energy models to calculate $d_A$.  The others used anchored BAO, where the standard BAO ruler was calibrated using CMB data together with an early universe cosmological model, including a number of neutrino families and neutrino masses, as well as a low redshift dark energy model.  Our treatment on the other hand has no assumptions regarding either the early universe cosmology of the evolution of the dark energy equation of state, although a calibration of $M_B(z)$ at one redshift ($z\sim 2.34)$ is only useful at another redshift ($z\sim 3$\ or $4$) if the function $M_B(z)$ is reasonably well-behaved, as is suggested for example by the linear dependence of peak luminosity on metallicity found in Ref.~\cite{timmes}.

\section{Remarks}

We have found no statistically significant evidence for the evolution of the absolute magnitudes of 1a SNe.  The uncertainty is dominated by the high redshift SN data, both the statistical fluctuations and the measurement precisions.  This is compounded by the fact that the redshift dependence of the shape, color and host corrections is unknown.  Furthermore, the hosts of some of the CANDELS SNe may well have been misidentified, leading to a significant error in $z$ which has not been considered in our study.  These limitations will largely be overcome by the James Webb Space Telescope (JWST), for example sufficient SN samples at a fixed redshift can provide the optimal corrections at that redshift.  Thus one may expect to achieve a much more stringent test of the evolution of 1a supernovae.

In particular, Ref.~\cite{dominguez} has estimated that evolutionary effects will be of order $0.20$, and so as the uncertainties fall well below $0.15$, in fact one may well expect that the JWST will discover evolution of the absolute magnitudes.  As we have seen in this paper, such a discovery will be independent of the cosmological model if these magnitudes are compared to BAO data.

At these redshifts, HETDEX and eBOSS will also have contributed firm measurements of the BAO scale, independent of those of BOSS.  For example, the forecast in Ref.~\cite{gongboeboss} shows that quasar-quasar correlations at eBOSS will be able to determine the line of sight and angular BAO scales at a number of redshifts.  This survey is already half complete, and the $1.8<z<2.0$ bin alone is expected to achieve a precision of 5.2\% and 7.4\% for $d_A$ and $d_H$ respectively.  Repeating the analysis in this note, interpolating the SN redshifts to $z=1.9$, this corresponds to a measurement of $M_B(1.9)-M_B(0.32)$ with a precision of $0.19$ which is entirely independent of the BOSS Lyman $\alpha$ BAO measurement.  At this lower redshift, one may also use SN at somewhat lower redshifts, driving the uncertainty down yet further.

\section* {Acknowledgement}

\noindent
JE is supported by NSFC grant 11375201. We thank Rados{\l}aw Wojtak for very insightful correspondence.


\end{document}